\documentclass{article}

\usepackage{PRIMEarxiv}
\usepackage[utf8]{inputenc} 
\usepackage[T1]{fontenc}    
\usepackage{hyperref}       
\usepackage{url}            
\usepackage{booktabs}       
\usepackage{amsfonts}       
\usepackage{amsmath}
\usepackage{nicefrac}       
\usepackage{microtype}      
\usepackage{lipsum}
\usepackage{fancyhdr}       
\usepackage{graphicx}       
\graphicspath{{media/}}     

\title{Semantic Segmentation of Porosity in 4D Spatio-Temporal X-ray µCT of Titanium Coated Ni wires using Deep Learning
}
\author{Pradyumna Elavarthi$^{1}$,
        Arun Bhattacharjee$^{2}$, Anca Ralescu$^{1}$, Ashley Paz y Puente $^{2}$
\thanks{$^1$ Department of Electrical Engineering and Computer Science, University of Cincinnati, OH 45221 USA.}%
\thanks{$^2$ Department of Mechanical and Materials Science, University of Cincinnati, OH 45221 USA.}%
}

\begin{document}
\maketitle

\begin{abstract}
A fully convolutional neural network was used to measure the evolution of the volume fraction of two different Kirkendall pores during the homogenization of Ti-coated Ni wires. Traditional methods like Otsu’s thresholding and the Largest connected component analysis were used to obtain the masks for training the segmentation model. Once trained, the model was used to semantically segment the two types of pores at different stages in their evolution. Masks of the pores predicted by the network were then used to measure the volume fraction of porosity at $0 mins$, $240 mins$, and $480 mins$ of homogenization. The model predicted an increase in porosity for one type of pore and a decrease in porosity for another type of pore due to pore sintering, and it achieved an F1 Score of $0.95$.
\end{abstract}


\section{Introduction}
In the field of metallurgy, it is highly important to know the evolution of microstructure to understand how a pure metal or alloy obtains certain mechanical properties. Traditionally, pathways for such micro-structural evolution have been predicted and understood based on physical and chemical principles. However, until such evolution is verified by direct observation a certain degree of uncertainty inevitably lingers regarding the exact pathway by which such micro-structural evolution occurs. To address this issue, \emph{in-situ} material characterization techniques have been extensively developed over the past few decades. Among these, synchrotron-based characterization techniques have received considerable attention across multiple fields of research ranging from mineralogy to biomedical sciences.\par
X-ray computed tomography is one such synchrotron-based technique that allows for the analysis of different phases and porosity present in metallic and mineral samples. The technique uses a computer algorithm to reconstruct the 3D volume of a sample from several 2D radiographs obtained at various angles while rotating the sample over 180◦ or 360◦. rotation. Depending on the detection mode of either transmission or absorption, different information about the microstructure of the material and its properties such as porosity and inter-metallic phases can be obtained. Once the 3D volume is reconstructed it can be uniformly sliced into multiple 2D reconstructed images representing the microstructure of the sample at different locations. With the incorporation of induction furnaces in the beamlines, \emph{in-situ} imaging of samples has become possible such that the exact microstructure evolution of a sample can be detected in real-time during different types of heat treatment.\par
On the analysis side, the large volume of data obtained from X-ray CT characterization presents a clear challenge to be individually analyzed by researchers. Conventional image segmentation techniques like Otsu’s thresholding\cite{one} are insufficient because of the inconsistencies in the shapes and intensities of the pores. To solve this problem there has been a growing interest in the use of machine learning algorithms to detect specific microstructural features from similar images\cite{two, three, four}. However, this requires training a model, in this work a convolutional neural network with masks representing the feature to be detected \cite{four} segmented the phases present in x-ray CT images of rock samples. A recent study performed by Gobert et.al\cite{two} on the x-ray CT data of 3D printed Al $12$ wt.\% Si shows that it is possible to detect porosity in metallic samples with a high degree of accuracy using machine learning. Badran et.al\cite{five} demonstrated the higher accuracy of segmenting SiC fibers from SiC matrix in CT data using synthetic images compared to manually segmented images where both phases have similar intensity values[5].The next step in this process would be to develop and train a network to detect and classify different pores present in the sample image. A major challenge to the development of such a network is the fact that the regions containing similar phases or porosity have the same pixel intensity and the shape and size of the pores evolve during the homogenization of the sample. Since phase segmentation and pore segmentation are usually dependent on the difference in the intensity of the pixels present, standard intensity-based thresholding techniques cannot segment and classify two different pores or phases present in the sample.\par
Segmentation and classification of such pores are of vital importance to the field of metallurgy and material science in general, as pores in the sample cross-section image can have different origins and it is important to classify them accordingly in order to analyze the properties of the samples. Furthermore, during \emph{in-situ} experiments different pores can have different evolutions even when undergoing the same phenomenon. An example of such behavior is the evolution of pores during densification when metallic or ceramic powders are sintered.\par
\begin{figure}
\centering
\includegraphics[width= 9cm]{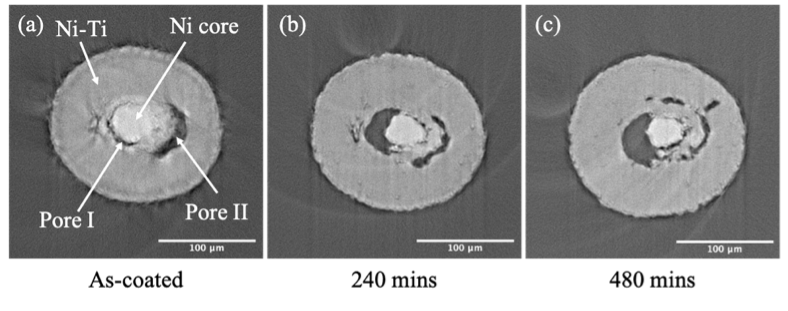}
    \caption{2D reconstructed tomographs of pack titanized 100 µm initial diameter Ni wires at (a) as-coated, (b) 240 mins and (c) 480 mins.}
    \label{fig:figure1}
\end{figure}

In our previous work, we performed \emph{in-situ} x-ray tomography on pack titanized Ni wires during homogenization. The 2D reconstructed tomograms revealed the presence of a dual pore structure in $25$, $75$ and $100$ µm samples with one of the pores forming at the center of the sample adjacent to the Ni core and the other inside the Ni-Ti region as shown in Figure \ref{fig:figure1}. During homogenization, the pore adjacent to the Ni core labeled as pore $I$ grows in size whereas the other crescent-shaped off-center pore, labeled as pore II, sinters. A detailed description of the experimental procedure can be found in \cite{six}.\par

Deep learning models have been very powerful tools in solving various computer vision tasks like image classification, segmentation, and object detection. In many instances, when there is abundant training data, convolutional neural networks are superior in performance to the traditional computer vision algorithms\cite{seven}. Deep learning models have remarkable generalization\cite{eight} ability and have been extensively used across numerous domains including segmenting regions of interest in x-ray tomograms in the field of medical science. A popular deep learning architecture called U-net\cite{nine} initially proposed for cell histology segmentation from CT images has since been used widely in bio-medical image segmentation tasks. These powerful algorithms, however, rely on excessive training data for learning the representations used for segmentation. In this paper, we used a fully convolutional neural network for the semantic segmentation of two different types of pores in the tomography data. To obtain the training data, multiple computer vision algorithms with varying parameters were used. These masks we then used to train the model based on the architecture of U-net to automate the process for the entire stack.\par
\section{Experimental}
\subsection{Mask Preparation}
\begin{figure*}[h]
\centering
\includegraphics[width= 10cm]{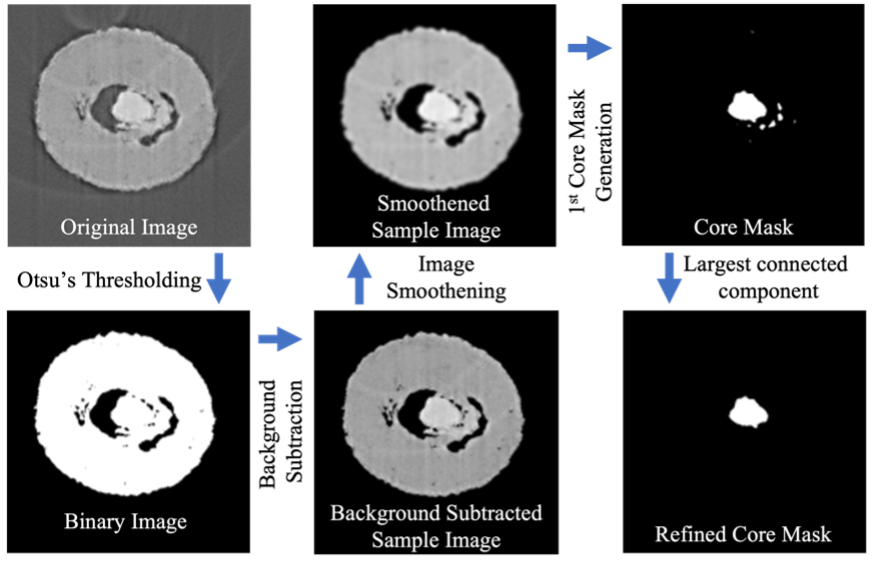}
    \caption{Schematic representation of the Ni core extraction algorithm.}
    \label{fig:figure2}
\end{figure*}
The dataset consists of 3 three image stacks acquired at $0$mins, $240$mins, and $480$mins during in-situ homogenization. Each stack consists of $2160$ images.The images have a very high resolution of $4096 \times 4096$ pixels. However, the sample occupies a very small region in the image and the rest is comprised of useless pixels pertaining to the sample holder and the background.  Computing the whole image was deemed unnecessary as it greatly increases the computation time and increases the risk of overfitting by providing more information about the background rather than the sample itself, thereby enabling the deep learning model to exploit that information \cite{ten, eleven}. Hence, the region of interest was separated for each image in the entire stack for each of the three datasets. After cropping, each image contains the pixels that belong to the sample and a few background pixels as shown in figure \ref{fig:figure1}. The background pixels and porosity of the sample have the same intensity as they both represent the lack of sample pixels. So, to isolate the sample from the background noise, the image was first smoothened using $3 \times 3$ Gaussian filter followed by Otsu’s thresholding method to create a binary image of the sample. Otsu’s method works by selecting the threshold pixel intensity that minimizes the intra-class variance and maximizes the inter-class variance \cite{one}. The binary mask was then inverted and subtracted from the original image to get the image of the sample without the background. The background-subtracted image was smoothened with a $3 \times 3$ Gaussian filter. The smoothened resulting image was then converted to binary using Otsu’s method to extract the nickel core. Since the intensity of the nickel core is almost similar to the intensity of the sample and because the size is significantly smaller than the sample, only $30$\% of the total non-zero pixels were selected in the nondecreasing order of their intensities for calculating the thresholding intensity using Otsu’s method. The resulting binary image is found to contain a portion of the sample surrounding the nickel core which could be explained by the formation of the intermetallic phase Ni$_{3}$Ti, which has a similar intensity range to Nickel. Since the core is always disconnected from the surrounding Ni$_{3}$Ti, the connected component analysis was used to get the core which is the largest strongly connected component. The connected component analysis is an algorithmic application of graph theory to find the connectivity of island-like regions in a binary image. The region with the largest connectivity is considered the largest connected component. The schematic of the above process is shown in figure \ref{fig:figure2} \par
The nickel core was then subtracted from the background-subtracted image to create a void in the core region which resembles porosity. This step is later used in the process to differentiate between the pores. Otsu’s method was then used to create a binary mask of the sample without the core. The mask was then inverted, and the background was eliminated by removing the largest strongly connected component from the inverted image to get the mask of the two pores. The mask of the two pores can be acquired without performing the core separation. Since the intensity of the two pores is significantly lesser than the sample’s intensity, performing Otsu’s thresholding would suffice to output the masks of the two pores together. However, the subsequent steps for separating the two pores in this mask are extremely difficult owing to the similar properties of both regions. For example, it is not necessary that the centroids of the pores always follow a specific geometric distribution that could be used to separate the pores. Since the pores grow and sinter over time \cite{six}, even the size property of the region is not consistent to be used for segmenting the pores.\par
\begin{figure*}[h]
\centering
\includegraphics[width= 16cm]{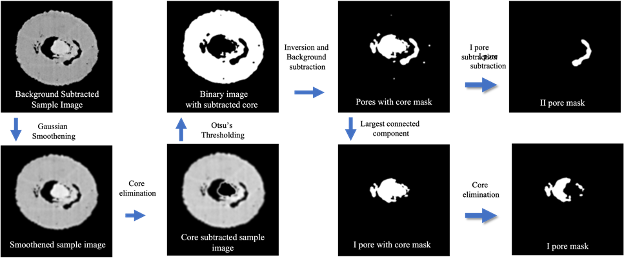}
    \caption{Schematic representation of the pore I and II generation algorithm.}
    \label{fig:figure3}
\end{figure*}

In the metallurgical sense, the first pore is a vacant region sharing an interface with the core. Hence, the classification can be done by using the fact that the pore that shares the boundary with the core is the central pore. Since the central pore is always connected to the core, eliminating the core from the image associates the area occupied by the core with the central pore region. The resultant area of the new region is the largest connected component since the area of the nickel core together with the area of the first pore would always be greater than the area of the second pore. So, the largest connected component is extracted as the 1st pore mask and is then subtracted from the pores mask to get the second pore. The core is then added to the first pore mask to get the actual first pore region. The final mask is created by concatenating pore I, pore II, the rest of the sample, and the background pixels since each pixel must be assigned a class label. The schematic of the process is shown in figure 3\ref{fig:figure3}\par

\subsection{Model}
\begin{figure}[!h]
\centering
\includegraphics[width= 14cm]{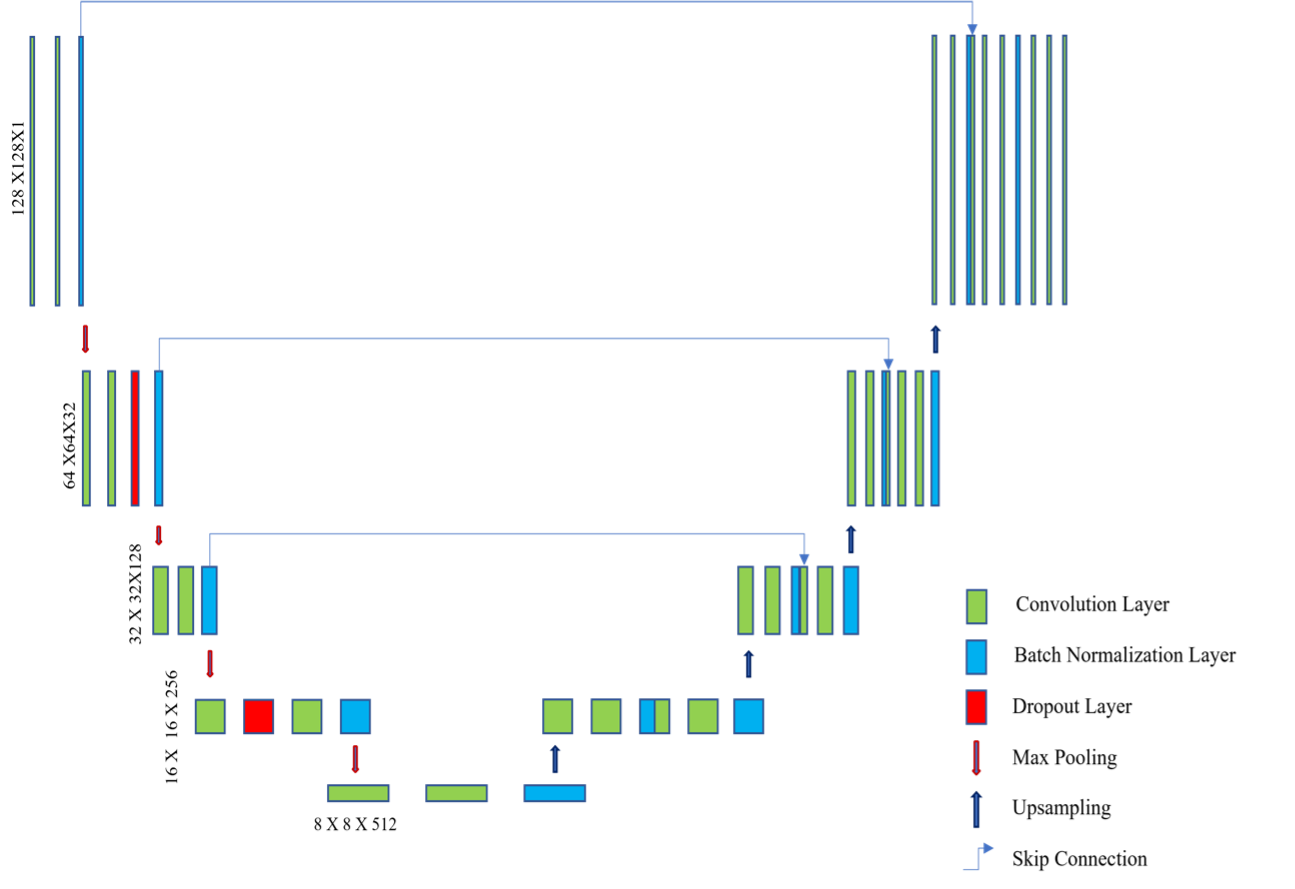}
    \caption{ Schematic of fully convolutional network used in the experiment}\label{fig:figure4}
\end{figure}

The network used in this paper follows the architecture of U-Net, designed for segmenting bio-medical images. The schematic of the architecture is shown in figure \ref{fig:figure4}. It is based on the Fully Convolutional Neural Network architecture and uses extensive image augmentation to compensate for the lack of abundant data. In the network, the image is first downsized in multiple steps to extract the relevant features and then upsized to the initial size outputting the mask. This architecture is selected because the tomography images have a high degree of similarity to the medical x-ray CT images. The images were first resized to $128 \times 128$ using bi-linear interpolation in OpenCV. The resized images were normalized to have a mean of $0$ and a standard deviation of $1$. The architecture of the model consists of encoder and decoder networks. The encoder network follows a typical classification CNN architecture. The encoder downsizes an input image from $128 \times 128$ pixels to $8 \times 8$ pixels using $3 \times 3$ and $1 \times 1$ convolutions and max pooling layers. It consists of repeated $3 \times 3$ convolutional layers followed by ReLU activation. ReLU activation was used because it is an unbounded function so effective against the vanishing gradient problem and has faster computational time since it does not activate the negative values creating a sparse output. After every two convolutions, a Dropout layer with a keep probability of $0.25$ is used for regularization. This layer probabilistically drops the input nodes to provide effective regularization. The output was then batch normalized. The down-sampling of the image was done by using max pooling with $2 \times 2$ filter with stride $2$. The image was downsized to $8 \times 8$ pixels and passed on to the decoder network. The decoder network is made with transposed convolutions to upsample the image to $128 \times 128$. In every step, the feature maps were doubled, and the number of channels was reduced by half. The feature maps in the decoder network were concatenated with the corresponding layers in the encoder network to preserve the spatial information lost during down-sampling. The final layer of the network is a $1 \times 1$ convolution to map $32$- dimensional feature vector to four classes. A total of $28$ convolutions are used in the model. The final layer uses Softmax as the activation function, which takes a vector of dimension $z$ and normalizes it to probability distribution proportional to the exponents of $z$ real numbers ($z$ being the number of output classes).\par
\subsection{Model Training}
Training a model refers to teaching a deep learning model with the data and the ground truth labels to update the trainable network parameters to learn the important features in the data. Similarly, testing refers to evaluating the deep learning model’s prediction of the data it has not seen before in training. The testing accuracy is used as a metric to evaluate the generalizability of the deep learning model.\par
The dataset used for training consists of 460 images carefully sampled from $0$ mins, $240$ mins, and $480mins$ images to provide the ground truth: $70$\% of the images were used in training, $30$\% were used in testing and validation. The training data was then augmented using domain-specific geometric image transformation methods such as flipping, zooming, rotating, and shifting. The data was augmented to increase the size of the training set and to improve the model’s ability to generalize. Augmenting the data using rotations enables the model to learn orientation invariant features\cite{twelve}. Since the data acquisition for tomography is variable and has a possibility of image artifacts and a chance of image noise, it is important to make the model robust to these small perturbations. Hence for effective generalization, gaussian noise with varying mean and variance was used as a data augmentation technique as shown in figure 5. Adding random noise to the image makes it ‘new’ each time the model processes it thereby making the model less likely to memorize the inputs and exploit the non-causal features present in training data\cite{thirteen}. Injecting noise as a form of data augmentation has a similar effect to regularization\cite{fourteen}.
\begin{figure}[!h]
\centering
\includegraphics[width= 9cm]{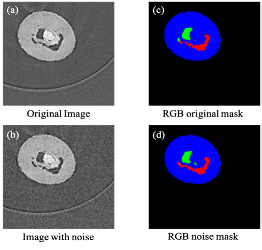}
    \caption{(a) original 2D reconstructed tomographic image, (b) 2D reconstructed tomograph with gaussian noise, (c) RBG mask for the original image,  and (d) RGB mask for the image with gaussian noise  showing the effect of gaussian noise addition .}\label{fig:fig5}
\end{figure}
The model is trained for 100 epochs with a batch size of 16 and using Adam and SGD optimizers. Adam is an efficient optimizer for gradient-based optimization of stochastic objective functions[15]. The model is trained on NVIDIA Geforce 1070 GPU. The training is continued for 50 epochs and weights of the epoch with the best loss are saved. Categorical cross-entropy in Keras was used as the loss function as the output has multiple classes. Cross entropy is a model performance measure that calculates the total entropy between two probability distributions. The loss function equation without the regularization terms is\par
\begin{align}
f(x) = -\sum_{i =1}^{n} \left(a_i*log(p_i)\right)
\end{align}
Where $a_i$ is the truth label, $p_i$ is the probability of prediction and n is the number of classes. The model was evaluated with the performance metrics like Precision, Recall, and F1 measure, as shown in table \ref{table:metrics}. Precision is the fraction of relevant predicted pixels among the predicted pixels whereas Recall is the fraction of relevant pixels in the predicted pixels. In other words, Precision measures how trustworthy the model is, and Recall measures how well the model can detect that particular class. F1 measure is the harmonic mean of Precision and Recall
\begin{align}
precision = \frac{Tp}{Tp + Fp}
\end{align}
\begin{align}
recall = \frac{Tp}{Tp+Fn}
\end{align}
\begin{align}
\textit{F1 Score} = \frac{2*precision*recall}{precision + recall} = \frac{2*Tp}{2*Tp + Fn + Fp}
\end{align}
where $T_p$ represents the True positives and $F_n$ represents the False Negatives. $F1$ Score is used because it can combine Precision and Recall into one measure that has both properties.
The resultant masks from the network are used to calculate the percent porosity of each pore. The total number of white pixels in each mask is considered the area of the pore. Using a complete stack of 2D images the percent porosity is defined as
\par
\begin{align}
\textit{Percent Porosity} = \frac{N_p}{Ns}\times100
\end{align}
where $N_p$ represents the total number of pixels occupied by the pore and Ns represents the total number of pixels occupied by the sample.\par

\section{Results and Discussion}
\subsection{Intra-sample intensity variation}
\begin{figure}[!ht]
\centering
\includegraphics[height = 7cm,width= 10cm]{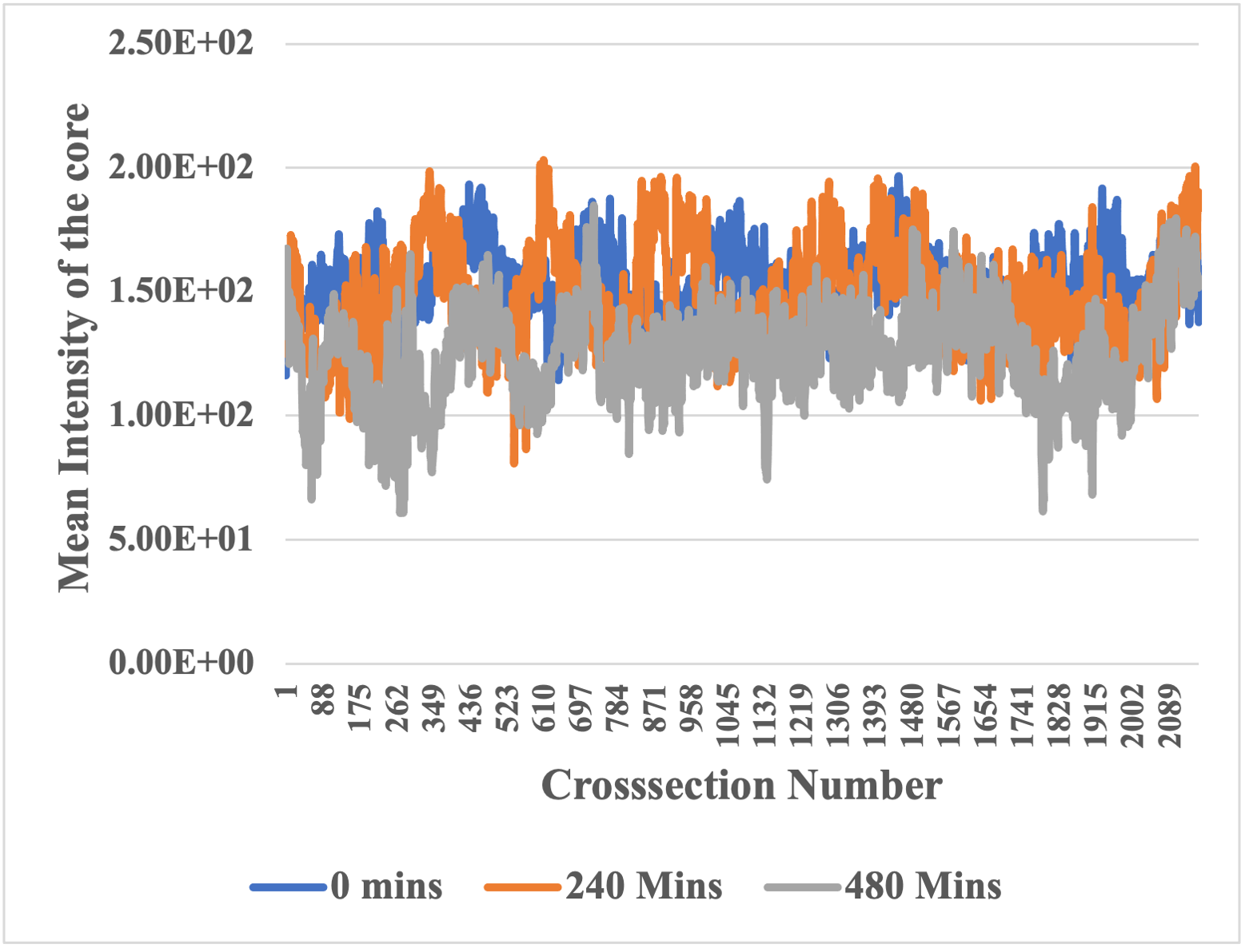}
    \caption{Intensity plot showing the variance in core intensity for the sample at (a) 0 mins, (b) 240 mins and (c) 480 mins.}\label{fig:figure6}
\end{figure}

To extract the nickel core from an image in the first dataset, the binary threshold intensity value was assumed to be constant for the entire stack. However, using one intensity value for the entire stack resulted in irrelevant pixels being extracted for some samples. The best results were observed when the threshold limit was calculated using Otsu’s technique on the top $30$\% of the non-zero pixels. The intra-stack variance in the intensity of the nickel core could be because of minor variance in the beam conditions over the total experimental time of $8$ hrs. The variance in the mean pore intensity across each dataset is shown in figure \ref{fig:figure6}. In the final step of mask preparation, consecutive masks of pore I, pore II, crust, and background are concatenated. While concatenating the masks it is observed that one pixel can be assigned to multiple classes since each mask is independently created. This problem was solved by subtracting the masks from each other so that the possible overlap of the pixels could be eliminated. Since the deep learning model is trained using categorical cross entropy each pixel should be assigned to one class only as otherwise, the model might fail to converge.
\subsection{Segmentation model results}
\begin{figure}[!ht]
\centering
\includegraphics[height = 6cm,width= 9cm]{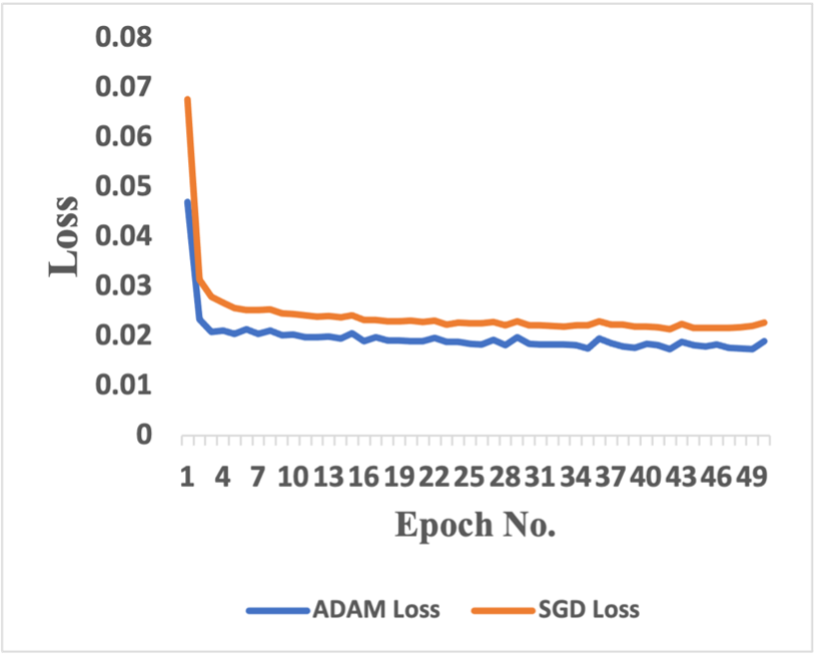}
    \caption{Training Loss curve for Adam and SGD}\label{fig:figure7}
\end{figure}

The learning curves for the models trained using Adam and SGD optimizers are shown in figure \ref{fig:figure7}. It can be seen that the model using Adam optimizer converged faster than the model using SGD because of the higher learning rate. Generally, Adam converges faster than the other optimizers, however, SGD is sometimes shown to have better generalizability than Adam\cite{sixteen, seventeen, eighteen}.
\begin{table}
\caption{Model Evaluation Metrics}
\centering
\begin{tabular}{ |c|c|c|c| } 
 \hline
        & Precision & Recall & F1-Score \\ 
         \hline
 Pore I & $0.97$ & $0.93$ & $0.95$ \\ 
  \hline
 Pore II & $0.98$ & $0.93$ & $0.95$ \\ 
 
\end{tabular}
 \label{table:metrics}
\end{table}

Precision, recall, and F1-Score are calculated for pore I and pore II. The test results on the three datasets are shown in table \ref{table:metrics}. The network is validated with the masks created in step 1. All the above-mentioned performance metrics were calculated using the created masks as the ground truth. However, by visual inspection, the masks generated by the model are better than the masks used for training the model since the smoothing functions used in the mask preparation process made the pore appear slightly bigger or smaller than the actual size to increase the variety. The performance of the network on the test data shows that this network generalizes well for the dual pore tomography data. The weights of this model can be used to speed up the training on similar datasets. For data similar to that used in this work, the model can be used to automate porosity detection and measurement with lower computation time.\par
\subsection{Visualization of the pore structure}
\begin{figure}[!ht]
\centering
\includegraphics[width= 9cm]{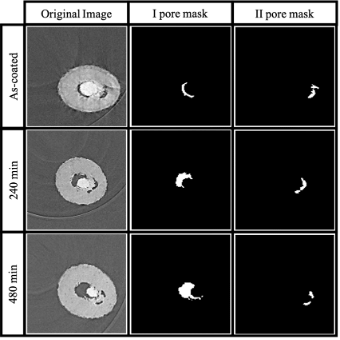}
    \caption{Predicted binary masks for pore I and pore II for a representative sample cross-section at (a) 0 mins, (b) 240 mins and (c) 480 mins.}\label{fig:figure8}
\end{figure}
The masks predicted by the network for pore I and pore II were separately visualized using TomViz software for the three different time steps. Measurement of porosity was performed separately for the two types of pores using their respective masks for the entire stack of 2160 images for three different time steps.
\begin{figure}[!ht]
\centering
\includegraphics[width= 9cm]{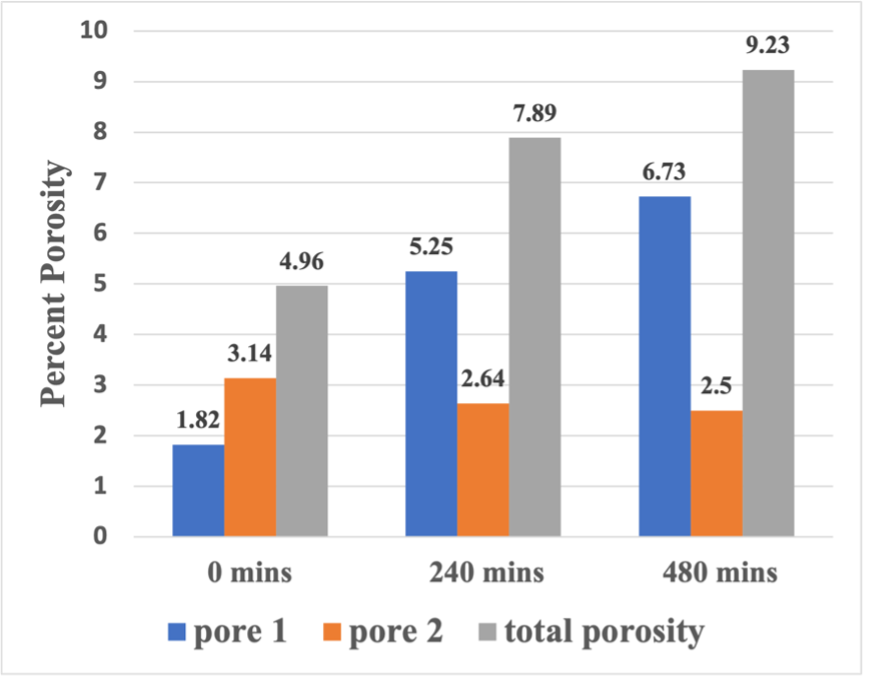}
    \caption{Percentage porosity of each pore in all the datasets}\label{fig:figure9}
\end{figure}
\subsubsection{Pore structure for Pore I}
\begin{figure}[!ht]
\centering
\includegraphics[width= 9cm]{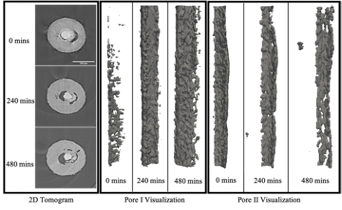}
    \caption{Visualization of pores I and II based on the binary masks predicted by the network.}\label{fig: figure10}
\end{figure}

Pore I appears as the smaller pore in the as-coated cross-section of the sample as shown in\ref{fig: figure10}. Visualization of the pore structure using the predicted masks shows the pore structure to be fragmented and disconnected. However, after 240 mins of homogenization significant growth in porosity occurs which exhibits a uniformly formed microtube through the sample as shown in figure \ref{fig:figure9}. Further increase in porosity occurs after $480$ mins of homogenization and the microtube size grows as the Ni core gets consumed. The increase in porosity during homogenization is due to the coalescence of Kirkendall pores.
\subsubsection{Pore Structure of Pore II}
Pore II is fully formed in the as-coated cross-section of the sample. After $240$ mins of homogenization, the volume fraction of pore II decreases significantly and its pore structure no longer appears to be uniformly formed. Further reduction in the volume fraction of the pore II appears to take place after $480$ mins. The reduction of the volume fraction of pore II can be attributed to sintering as pore II is completely inside the NiTi phase in the sample.
\subsubsection{Effect of Pore I and II on overall porosity}
The total porosity of the sample was found to be constantly increasing during homogenization. Although the increase in porosity after $240$ mins of homogenization was found to be higher than between $240$ mins and $480$ mins. However, the increase in the total porosity was always smaller than the increase in the porosity of pore I. This is due to the fact that the porosity of pore II decreased during homogenization. 

\section{Conclusion}
This work demonstrated an approach to create and extract masks required for training a deep learning segmentation model using adaptive thresholding and connected component analysis. These masks were then used to train a fully convolutional neural network based on the architecture of the U-Net segmentation model to semantically segment different porosity in the sample. The training images are carefully selected by visual inspection to make sure that masks with errors are not fed to the network. To improve the robustness of the model towards distortions and image artifacts, domain-specific data augmentation was used during training. Random Gaussian noise with varying mean and variance was also used as a form of data augmentation technique to improve the model’s reliability on causal relationships. The model trained with these settings achieved an F1-Score of $0.95$ on the test images. 
Our analysis of the results predicted by the model showed that porosity increase after the first $240$ mins of homogenization was found to be higher than in the subsequent  $240$ mins. However, the rate of increase in the total porosity was always smaller than the rate of increase in the porosity of pore I, because the porosity of pore II decreased during homogenization in the dual pore structure exhibited by Ti-coated Ni wires in x-ray tomography images.

\bibliographystyle{unsrt}  
\bibliography{references}

\end{document}